# KNOWLEDGE MANAGEMENT CONCEPTS FOR TRAINING BY PROJECT
*An observation of the case of project management education*

Christine Michel, Patrick Prévôt
*Laboratory LIESP*
*Univeristy of Lyon, INSA-Lyon,*
*Bâtiment Léonard de Vinci  21, avenue Jean Capelle*
*F-69621, France*
Christine.Michel@insa-lyon.fr, Patrick.Prevot@insa-lyon.fr

Keywords: KM, expanded learning circle, teaching by project, project management education, tutor activity

Abstract: Project management education programmes are often proposed in higher education to give students competences in project planning (Gantt's chart), project organizing, human and technical resource management, quality control and also social competences (collaboration, communication), emotional ones (empathy, consideration of the other, humour, ethics), and organizational ones (leadership, political vision, and so on). This training is often given according a training-by-project type of learning with case studies. This article presents one course characterized by a pedagogical organization based upon Knowledge Management (KM) concepts: knowledge transfer and construction throughout a learning circle and social interactions. The course is supported by a rich and complex tutor organization. We have observed this course by using another KM method inspired from KADS with various return of experience formalized into cards and charts. Our intention is, according to the model of Argyris and Schön (Smith, 2001), to gain feedback information about local and global processes and about actors' experience in order to improve the course. This paper describes precisely the course (pedagogical method and tutor activity) and the KM observation method permitting to identify problem to solve. In our case, we observe problem of pedagogical coordination and skills acquisition. We propose to design a metacognitive tool for tutors and students, usable for improving knowledge construction and learning process organisation.

## 1 INTRODUCTION: FORMATION WITH PROJECT MANAGEMENT

Training in project management is growing significantly in higher education, particularly in engineering schools and postgraduate schools. Indeed, the study of Thomas & Mengel (2008) on the evolution of this discipline in higher education shows than between 2004 and 2007 the number of programmes concerned with project management education increased from 6982 to 12500 (an increase of 79%). Training in project management education, on the other hand, did not significantly change during this period, in spite of the recommendations and suggestions of the Project Management Institute (PMI). The study of Thomas and Mengel shows that training in project management education must take into account, at the same time, 'hard' and 'soft' competences. Hard competences correspond to knowing how to plan the project (Gantt's chart), to organize the project management, to manage resources, to control quality, to handle follow-up and closure (receipt) of the project, to use tools for automation with mature technology, to include/understand and formalize the customer requirements, to organize the reporting of the project, and to learn from its errors or good practice (Manzil-e-Maqsood & Javed, 2007). Soft competences correspond to social competences (collaboration, communication), emotional ones (empathy, consideration of the other, humour, ethics), and organizational ones (leadership, political vision, and so on) (Thomas & Mengel, 2008; Berggren & Söderlund, 2008; Crawford et al., 2006).

The type of learning best adapted to this education is training–by-project (Bredillet, 2008). The creation process of knowledge and competences is based upon social interaction and direct experimentation. It substitutes to traditional type of learning a dynamic of co-development, collective responsibility and co-operation (Huber 2005). The learner is an actor and the principal author of his/her learning. A significant enrichment arises from his/her activity, both for him/her and all the other learners. A consequence of this training is to segment the class into sub-grouped projects, driven by tutors. However, the coordination and harmonization of their activities is extremely difficult to realize when each group functions autonomously, on different subjects or in real and varied environments (for example, enterprises) and when, moreover, the project is conducted over long periods (more than four weeks). Moreover, these contexts make the perception of individual and group activity difficult, especially if no technical support regarding information and communication is used.

This article presents a project management training course characterized by a pedagogical organisation based upon training-by-project with Knowledge Management (KM) concept. Indeed, knowledge transfer and construction is made according to a learning circle and social interactions. As this is explained further, the course is supported by a rich and complex tutor organization. In order to analyze this formation, we have made an observation through another KM method. Our intention is, according to the model of Argyris and Schön (Smith, 2001), to gain feedback information about local and global processes and about actors' experience in order to improve the course. This paper describes precisely the course (pedagogical method and tutor activity) and the KM observation method permitting to identify problem to solve. In our case, we observe problem of pedacogical coordination and skills acquisition. We propose to design a metacognitive tool for tutors and students, usable for improving knowledge construction and learning process organisation.

## 2 ANALYSIS OF THE PROJECT MANAGEMENT COURSE

### 2.1. Organization of the course

The course is composed of a theoretical presentation on the principles and methods of project management and their practical application to a project (called 'PCo' for 'collective project') carried out in groups (12 groups of eight students working on different industrial needs). Envisaged by Patrick Prévôt (Prévôt, 2008), the project management course lasts six months and corresponds to an investment of approximately 3000 student working hours per project. The teaching objectives (Dpt GI, 2008) are to acquire hard and soft competences previously cited. The teaching team is composed by 24 tutors (2 tutors technical and management per project group), two managers (technical and management) charged to coordinate the activities of respectively technical and management tutors, a teacher presenting theoretical concepts and a coordinator/director responsible for the organization of the learning and training of all the groups.

The project is structured in four phases (Perrier, 2008). (1) November : *response to the call for tender* (formalization of the client's requirements.) (2) December : *master plan* (means, tools and organization of the team project) definition of tools to drive the project (*dashboard*) and rules to test the quality of deliverables (*rules of receipt*). (3) January to March : *production* (of a product or a study). (4) Until mid-April : *deliverable of project closure writing* : *technical report* which describes the product and *management report* which is an analysis, from the student's point of view, of the flow and problems of the project. The project is closed by one *dramatized presentation* in front of all the actors of the project

### 2.2. Teaching model

This formation is specifically interesting according the KM research point of view because the experiential learning model used is closely linked to the Kolb circle (Cortez et al., 2008) or the Nonaka SECI circle (Nonaka et al., 2000) which explain the dynamic construction of the knowledge . Training is organized according to a loop of concrete experience, reflective observation, abstract conceptualization, presentation and active experimentation. Knowledge is acquired via active steps by the learner. Socio-constructive approaches add the use of the personal and social construction experiment.

More precisely in our course, the type of learning is organized according to Berggren's expanded learning circle (see Figure 1) which adds to the usual experiential model two concepts:

reflection/articulation and enaction (Berggren & Söderlund 2008).

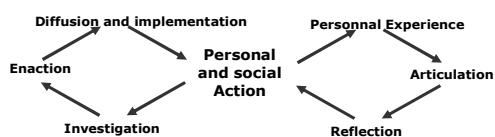

Fig. 1. Expanded learning circle

The *experience* of the student is a result of the education process constructed by following the small right circle or the large eight-form expanded circle. The *articulation* phase corresponds to debriefing discussion and debate driven by the tutor (one face-to-face discussion per week) or student project manager. The theoretical concepts of the courses, for example, are discussed, analyzed and understood through the reality of the project and the technical tutors take a large part in animating these phases. Conversely, the management tutors present and discuss with team the 'soft' competences which are needed or already exist. This work is strongly linked with *reflections* phase (especially *reflective observation*s). Reflective observation are realized in some cases in a tacit way after these discussions with tutors or in a more formal way in the management report or through the deliverables of production. In this case, reflection is articulated with conceptualization actions and helps to realize the 'hard' and 'soft' competences constructed. This is also the occasion for each individual to express his/her personal experience. By combining reflective perception, reflections (helped with articulation actions) and previous experience, the student is able to understand and apply theoretical concepts in the real project by following the tutor's instruction concerning *personal and social actions useful for the project*. It helps to construct the student's experience regarding the teaching objectives. The student is able also to choose and define his/her personal and social actions and construct another unique experience not so well formalized by the teaching team.

Another characteristic of our teaching model is to promote *investigative* action on the one hand, and *enaction and diffusion* actions on the other. Indeed, most of the course in project management consists of realizing a well known project (case study). In our case, *investigative* action is emphasized by the fact that students have to solve a real industrial problem without a predefined solution. It gives more of a challenge and motivates knowledge construction. The enaction and diffusion process is realized in *dramatized representations*. They have the same objective of supporting the reflection and conceptualization needed for students to realize the experience they gain, but also take part in a KM diffusion process, between project team, teaching team and scholar department. Students present here their good and bad practices, and their feelings and judgements about the formation and the tutors.

This circle of 'experience, articulation, reflection, action' is the foundation of teaching-by-project type of learning which resolution of a problem is combined with socio-constructivism theories (as previously illustrated by Gibbons' point of view). It helps with the construction of knowledge and supports individual motivation. Indeed, the fact of having to confront different points of view helps the cognitive process and reinforces social motivation. Often, formations in project management education highlight the role of action and experimentation. As does Berggren (Berggren & Söderlund, 2008), we consider that articulation between *action* and *reflection* is also fundamental because it supports the **evolution of behaviour** and we propose to use it in other contexts like lifelong professional learning for example (Michel, 2008). More than 'trained technicians', we want to form 'reflective practitioners' as Crawford said (Crawford et al., 2006), who are able to choose how to learn and evolve according to future unknown contexts.

## 3 OBSERVATION OF THE PROJECT MANAGEMENT COURSE

### 3.1 Method of observation

The methodology used is adapted from MKSM model and KADS model (Dieng et al., 2005). These methods, starting from documents produced by an organization and talks with the actors, model complex industrial systems by identifying and inter-relating various concepts: product, actor, activity, rules and constraint. Each concept is defined on a card. The ICARE (information, constraint, activity, rule, entity) cards describe any object precisely intervening in the process. The RISE (reuse, improve and share experiment) cards describe any problem occurring during the process and specify the contexts, solutions suggested or recommendations. The elements described in the ICARE and RISE cards were organized overall in a *chart* which shows their interrelationships. The adaptation of methods MKSM and KADS to our context was carried out with the assistance of the director of Airbus's KM service (Toulouse), Rene Peltier.

The effective observation was carried out by a group of fifth year Industrial Engineering students as the framework of a KM course. The student used various sources of observation: *the formal documentation produced within the framework of PCo* (management report, tutor guides, rules of evaluation), *experience feedback* (return of experience called REX)) and *transfer of expertise* (expertise transfer called EXTRA) of the actors of the project. The REX were provided by the students themselves (they gave feedback on their own experience of the previous year when they were involved with the PCo) and by tutors. The EXTRA was provided with the director of the formation and tutors managers in order to formalize precisely their activities and responsibilities. The use of the student's REX was made directly by expressing their experiments in ICARE and RISE cards. This REX was made with 60 students (over two years) who had been involved with a PCo. The tutors' REX and EXTRA were done according to semi-directing talks directed and registered by students and then used to write ICARE and RISE cards. The information was provided by the director, six tutors and three students currently leading projects in the PCo formation.

### 3.2. Analysis of problems related to tutor activity

The problems experienced were expressed or described in 36 RISE cards. The majority relate to the management of the team work by the team itself and the teaching organization of the project. Nevertheless, many of cards mentioned problems concerning evaluation, presence, coherence and coordination of tutors. The students expressed a feeling of injustice concerning *the individual evaluation* because the notation is the same for all the members of the project (with about + or -2 points according to their investment), even if the students have invested themselves little or less than others. The tutors also universally expressed their impotence as regards being able to concretely evaluate the students individually. This impotence is explained by the intuitive and tacit character of the evaluations, by the lack of traceability of the students' actions, and by the absence of discussion with their colleagues. Some students underlined *deficit of communication* or missing *presence* of some tutors. Others mentioned deficit of *coherence*, *coordination* and *diffusion of information* concerning, for example, the instructions (which were described as ambiguous or contradictory) given to the various groups or concerning the way to practically apply the theoretical concepts.

If we observe more precisely the tutor activity we can said that this problem is not surprising. Indeed, the roles of the tutors are varied. Indeed, we describe in (Michel, 2009) and according to Garrot's taxonomy (Garrot, 2008) how tutors play various role like *social and intellectual catalyst*, *mediator*, *individualizer* or *autonomizer* for soft skills acquisition, or the role of *relational coach* for working in group and leadership skill, and also roles of *pedagogue, content expert, 'evaluator'* and *'qualimetror'* (i.e. quality measurer) for hard skills. Moreover the tutor number is large (24 tutors and 3 managers), and they have to work to a unique and non reproductible project. They work with student most of time in face-to-face and no organisation, communication or capitalization tool is used. For example, no specific tool for supervision is currently proposed to the tutors for the follow-up of student activities or their notation. The appreciation of student activity is done in an implicit way, according to the number and the quality of face-to-face student–tutor interactions. In terms of communication and coordination, each tutor works individually with his/her group and does not communicate systematically with his/her technical or management opposite in order to have a complete vision of the group activity.

(Billois *et al.*, 2009) have more specifically studied how to solve these problems with technical support based upon **dashboards**. Dashboards are devices of supervision built by the project leader in the second step of the PCo. It's a team supervision tool given to student various information (total working time, delays, etc.) used to help the progression of the project. We briefly present in the next section some results of this study.

### 3.3. Study of technically supported solutions

The study of the use of **dashboard** (Billois et al., 2009) showed that this tool is in fact little used. It exclusively remains a theoretical exercise, carried out by the leader of the project and almost never consulted by the other members of the project.

In order to improve the dashboard and the pedagogical process, Billois *et al* (2009) consider useful to add other communication and supervision function. Indeed, to complete the historical dashboard which corresponds to *Team feedback* supervision tool (highlight by a bold rectangle), he propose an evolution with the opening-up of a *student feedback* supervision tool presenting individual indicators (morale, working hours, and so on), the opening-up of *personal and team blog* and view of *tutor's schedule* to facilitate contact.

We models all the propositions in Fig 2. The mode of consultation (publication control/reading/writing) is represented by various arrows.

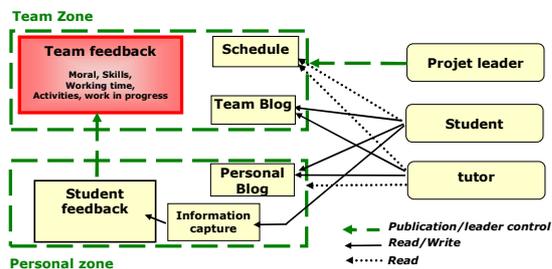

Fig. 2. Advanced dashboard

The evolutions were submitted to the tutors and to three project leaders currently in formation. Type of use and opening-up is considered as useful because it permit to student to train a management tool and so, by experiencing it, to better understand hard competences or soft ones (like team force for example). Spaces for expression regarding the project and each member were particularly appreciated both by the tutors and the project leaders. Indeed, the project leader give the example of the 'state of mind indicator' of the 'incident indicator' which gain to be contextualized so as to allow the tutor and the project leader to be aware of situations, to understand the reasons for the dysfunction and to initiate a dialogue with the members of the project. Reciprocally that makes it possible for the members to be aware of the state of mind of the project leader and thus to be more quickly given responsibilities. A space for communication intra-project does not seem desirable to the student whereas it is judged to be a good idea by the tutors. Concerning the historical dashboard (Team feedback), the actors are overall satisfied with the current type and form of the indicators and propose a small addition. The students wish, for example, to have the assistance of there tutors or project leader concerning with the skills needed for the project realization and strategies to develop them (like example for training or book to reading for example). In a more general way, they are asking for more discussion and formalism on this point. This would reassure them by providing a global vision about their mission and would allow better distribution of the training and activity for each member of the project.

## 4 DISCUSSION AND CONCLUSION

The studies of the project management course by KM method and especially the REX observation and the RISE cards show a lot of waiting concerning the support of tutors' activities. All actors are asking for new means of **coordination and collaboration** with the students and between tutors. Tutors express difficulty with **playing their teaching role**. The solution under consideration is to design an 'advanced' dashboard (supervision, communication and collaboration tool) adapted from the current ones. This solution would make provide a support for several of the tutors' roles, in particular those of *pedagogue, evaluator* and *'qualimetror' (i.e. quality measurer)*, by tracing the activities of the student and having direct access to their contributions. The use of the 'advanced' dashboard is particularly adapted to our teaching model, which is based on the expanded learning circle (Berggren & Söderlund, 2008) of Kolb (Cortez et al., 2008). The articulation between conceptualization and experimentation concerns various pedagogical tasks and activities and is usual in all project management training. The originality of our approach is to also consider articulation between action (experimentation or conceptualization) and reflexive practices. Indeed, like Berggren (Berggren & Söderlund, 2008) we think that project management education training highlights the role of experimentation and that it is necessary to balance teaching action with reflection. Indeed, this combination is well placed to accompany an **evolution of behaviour** in terms of skills (management, communication, collaboration and all 'soft' competences) and natural reaction (to be able to learn how to learn and evolve in surprising or unknown situations) by supporting the students' capacity to self-critically analyse. This capacity results mainly from the training activities carried out with the tutors and must be more supported. The dashboard proposition presented is good and must be useful but the indicators have to be rethought. Nevertheless, the results of observation and discussion with tutors and students have allowed us to realize that the dashboard takes on the role of a **metacognitive tool**. According to Azevedo (Azevedo, 2007) the term 'metacognitive tool' serves two goals: (1) acknowledging the role of metacognition in learning complex topics with CBLEs (computer based learning environments) and (2) extending the original classification of "computers as cognitive tools" by acknowledging the complexity of self-regulatory processes during learning with CBLEs. He states that the learners' self-regulatory processes may consider: *cognition*

(e.g., activating prior knowledge, planning, creating sub-goals, learning strategies), *metacognition* (e.g., the feeling of knowing, judgment of learning, content evaluation), *motivation* (e.g., self-efficacy, task value, interest, effort), or *behaviour* (e.g., engaging in help-seeking behaviour, modifying learning conditions, handling task difficulties and demands). Several of these concepts appear in the dashboard and need to be refined, such as *motivation* (which is global to the team) or *cognition* (which is currently represented very superficially by the 'skill acquired'). Others clearly are missing and correspond to tutors' or students' needs. We think, for example, of the need for judgements on the skills acquired, or the need for collaboration in the definition of training strategies which concern the fields of cognition and metacognition. Lastly, this type of tool can directly contribute to the realization of the management tutor's roles of '*meta-catalyst*', *individualizer* and *autonomizer*. Indeed, because it is impossible currently to automatically have the smoothness of perception of the tutor, concerning the social form of the team and the complexity of human psychologies, we think that the teaching activity of debriefing must continue to be carried out in face-to-face discussion, as is currently the case. Nevertheless the tutor must be able to be helped in his individual perception and his memorizing by technical supports like the content of personal and collective zones of expression and feedback.